# Ferroelectric and Hyper Dielectric modes in Ferronematic Liquid Crystals.

Rahul Uttam[1], Neelam Yadav[1], Alexander Belik[1], Wanhe Jiang[3], Georg H. Mehl[3], Jagdish K. Vij[1*], Yuri P. Panarin[1,2*]

[1] Department of Electronic and Electrical Engineering, Trinity College, Dublin 2, Ireland

[2] Department of Electrical and Electronic Engineering, TU Dublin, Dublin 7, Ireland

[3] Department of Chemistry, University of Hull, Hull HU6 7RX, UK

**Abstract:** Binary mixtures of the ferronematic compound DIO with recently reported non-ferroelectric material WJ-16 which shows Colossal Permittivity (CP) ~5000 and superparaelectricity (SPE) were studied by POM, electrical switching studies, and dielectric spectroscopy. Three mixtures with different contents of WJ-16 as 10, 25 and 50% in DIO as host were prepared. Our original expectation was the development of new nematic materials with both ferroelectric nematic (NF) and non-ferroelectric CP phases. The non-ferroelectric phase in mixtures exhibits a CP mode originally observed in pure WJ-16 and was termed as superparaelectric. However, the dielectric spectroscopy of mixtures shows two distinct relaxation processes: the typical paraelectric response and the CP mode. Therefore, this CP mode cannot be called superparaelectric and is redefined it as Hyper Dielectric mode. This is the first direct demonstration of materials with both ferroelectric and Hyper-dielectric phases in liquid crystalline materials. The hyper dielectric phase has a good potential as a working media for supercapacitors industry.

===========================================================

*****Corresponding authors:** Y. Panarin: yuri.panarin@tudublin, J. K. Vij: jvij@tcd.ie

**Keywords:** Ferroelectric nematics, dielectric spectroscopy, superparaelectric fluids, Colossal dielectric permittivity, ferroelectric.

**Highlights:**

- Binary Mixtures of ferroelectric nematic DIO and new hyper dielectric compound WJ-16 were prepared and studied.
- The resulting mixtures possess both ferroelectric and hyper-dielectric phases with Curie-Weiss type of transition between them.

- Both phases have giant dielectric permittivity which can be used in supercapacitors.

1. Introduction

The ferroelectric nematics were observed independently in two different nematic compounds, DIO [1] and RM-734 [2], and the discovery has generated great interest in the liquid crystal (LC) community. Ferronematic s compounds are characterized typically by (i) extremely large dipole moments ($\mu$ ~10 D), (ii) high spontaneous polarization ($P_s$ ~ 5 µC/cm$^2$) [3,4,5,6] and (iii) Colossal dielectric Permittivity (CP) ~ 10,000. Such high values for dielectric permittivity are modelled by two different theoretical models: the high-ε model [7,8] taking in account ferroelectric Goldstone (phason) mode, and the Polarization-External Capacitance Goldstone Reorientation model [9,10,11] taking in account coupling of the Goldstone mode with the external capacitance of the measuring cell, particularly through insulating surface layers at the electrodes.

Though new ferroelectric compounds [12,13,14,15,16,17,18,19,20]are e reported and discussed in the literature continuously, the discovery of ferroelectric nematics has not brought any benefit to the Electro-Optic (EO) switching and Liquid Crystal Display (LCD) industries. On the other hand, the huge or Colossal dielectric Permittivity (CP) observed in ferroelectric nematics can be utilized as a working medium for Supercapacitors. Supercapacitors are specially designed capacitors that have a very large capacitance. Supercapacitors combine the properties of capacitors and batteries into one device so that they can be used in two rather independent applications: (a) as batteries, i.e., storage for electrical energy; and (b) as electronic devices. Supercapacitor batteries are based on a range of storage mechanisms, such as the accumulation of ions in a double layer or electrochemical reactions of a combination of the ions. The specific electrical capacity of a typical supercapacitor is about 10 kW/kg, that is an order of magnitude higher than that of lithium-ion batteries and supercapacitors have a faster charge/discharge cycles [21]. This property is especially important in applications that require rapid bursts of energy to be released from the storage device. Supercapacitors can also be used as electronic devices in the electronics industry, where the need for increased device density goes hand in hand with the miniaturization of electronic structures that continue to operate efficiently when scaled down to the nanoscale. Thus, micro- and nano-electronic devices require the replacement of conventional dielectric materials with those that exhibit colossal permittivity, such as ferroelectric nematics. In addition to ferroelectric nematics and solid ferroelectrics, CP is also observed in solid non-ferroelectric materials known as superparaelectrics (SPEs).

There are no strict definitions of the terms ferro -, para -, and superpara-electrics since they were all introduced as electrical analogs of magnetic materials. Therefore, to avoid possible ambiguities and misunderstandings, we briefly dwell on their definitions. The modern history of ferroelectricity is more than 200 years old and is well summarized in the work by Lűker [22]. The first recorded observation of the pyroelectric effect is attributed to the Greek philosopher Theophrastus around 314 BC. He noted that the mineral Tourmaline (which he called "lyngourion") had the peculiar property of attracting light-weight materials like straw and bits of wood when heated. In 1824, David Brewster observed the phenomenon of pyroelectricity in various crystals, among which was Rochelle Salt, but perhaps the first systematic studies were those of Pierre and Paul-Jacques Curie in 1880, when they observed and described the piezoelectric effect. In 1912, Peter Debye suggested that this class of materials carry a permanent electric dipole moment in analogy to the ferromagnetics. Following Langevin's theory of paramagnetism, Debye suggested the equation $(\varepsilon-1)/(\varepsilon+2)=a+b/T$, where $a$ is proportional to the density of the substance and $b$ to the square of the electric dipole moment. According to his relation, for a critical temperature $T_C=b/(1-a)$, the value of the dielectric constant reaches infinity. Therefore, he proposed $T_C$ to be the analogue of the Curie temperature of a ferromagnet. Later, Schrödinger speculated, based on Debye's model, that all solids should become "ferroelektrisch" at a sufficiently low temperature. According to Schrödinger, the terms paraelectricity and ferroelectricity are two parts of the same coin, with the Curie temperature as the edge of the coin

Hence, following this *historical* definition, paraelectrics were introduced as the counterpart of ferroelectrics, and the paraelectric phase is highly non-linear with temperature and electric field. Taking into account that many materials are not ferroelectric, paraelectricity describes the behavior of dielectrics where an applied electric field causes polarization due to the alignment of dipoles parallel to the field, and this polarization disappears when the field is removed. Hence, for such materials, the *P-E* dependence is practically linear, i.e., it obeys the Langevin function, showing linear *P-E* in a sufficiently wide range of applied electric fields with a gradual saturation at higher (practically unreachable) fields. For example, our sample under study, WJ-16, shows linear *P-E* dependence even at strong electric fields up to 11 V/μm [23] and can be considered as linear for our experimental environments. On the other hand, this compound, being non-ferroelectric, exhibits CP and can be considered as a superparaelectric in analogy with superparamagnetics. In the strict sense of the word, only *linear dielectrics*, which consist of non-polar molecules, do show true linear *P-E* dependence for unlimited electric fields

In this paper, we show the results of the study of the mixtures of two different types of nematic LCs. These mixtures are the first example of LC materials with two different nematic phases- ferroelectric and hyper dielectric.

## 2. Materials

This paper reports a study mixture of two different types of nematic LCs: (i) well-known ferroelectric nematic DIO and (ii) newly synthesized non-ferroelectric, but superparaelectric nematic WJ-16 [23]. The molecular structure of WJ-16 is based on DIO, where a fluorophenyl group of DIO is replaced by a pyrimidine group, as shown in Fig. 1, and this replacement increases the molecular dipole moments from 9.4 D for DIO to 10.4 D for WJ-16. Surprisingly, WJ-16 does not show the ferroelectric nematic state but also shows CP. This was explained in the Ref. [24,25] which show that the high value of dipole moment is important but not sufficient for formation of $N_F$ phase.

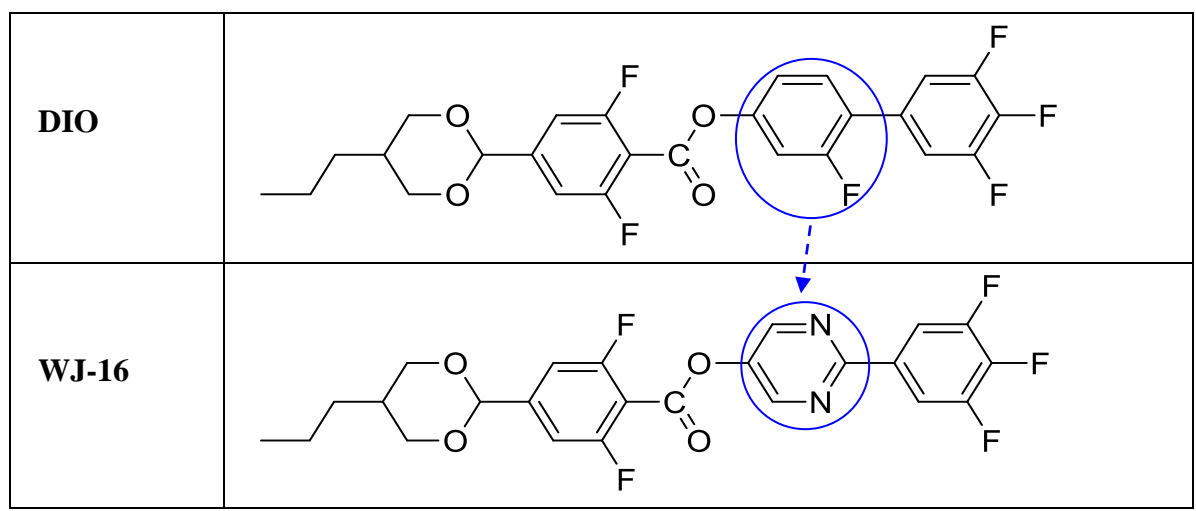

**Fig. 1.** The chemical structures of molecules WJ-16 and its prototype DIO

The transition temperatures of individual components and their mixtures are given in Table 1

| System | Phase Sequences |
| --- | --- |
| *WJ-16* | Cr (79.3) SmA (110.5) N (198.6) Iso |
| *DIO* | Cr (< 20) N$_F$ (66.8) N$_x$ (83.5) N (173.8) Iso |
| *MIX 10* | Cr (< 20) N$_F$ (52.0) N$_x$ (73.0) N (179.0) Iso |
| *MIX 25* | Cr (< 20) N$_F$ (50.0) N$_x$ (72.0) N (180.5) Iso |
| *MIX 50* | Cr (34.5) N$_F$ (47.5) N$_x$ (66.5) N (190.0) Iso |

**Table 1.** The phase sequences and phase transition temperatures (in ºC) for different Liquid Crystal systems and their mixtures.

From the first sight, the phase sequences of mixtures appear similar to DIO. However, in DIO, "N" is an ordinary nematic phase, but in WJ-16 and mixtures, "N" is superparaelectric nematic phase with the CP, as discussed below.

## 3. Results and discussions

*3.1 Polarizing Optical Microscopy (POM)*

The temperature-dependent optical textures of the prepared mixtures were investigated using polarizing optical microscopy (POM) on different cells with thicknesses ranging from 2 to 9 µm. Both commercially sourced (E.H.C. Co. Ltd Japan) and laboratory-fabricated "uncoated" indium tin oxide (ITO) electrode cells were used. Planar and homeotropic alignments were achieved by spin-coating appropriate alignment layers onto the ITO surfaces. Fig. 2 (a, b, and c) illustrates the POM textures obtained from 9 µm anti-parallel buffed planar cells filled with the mixtures

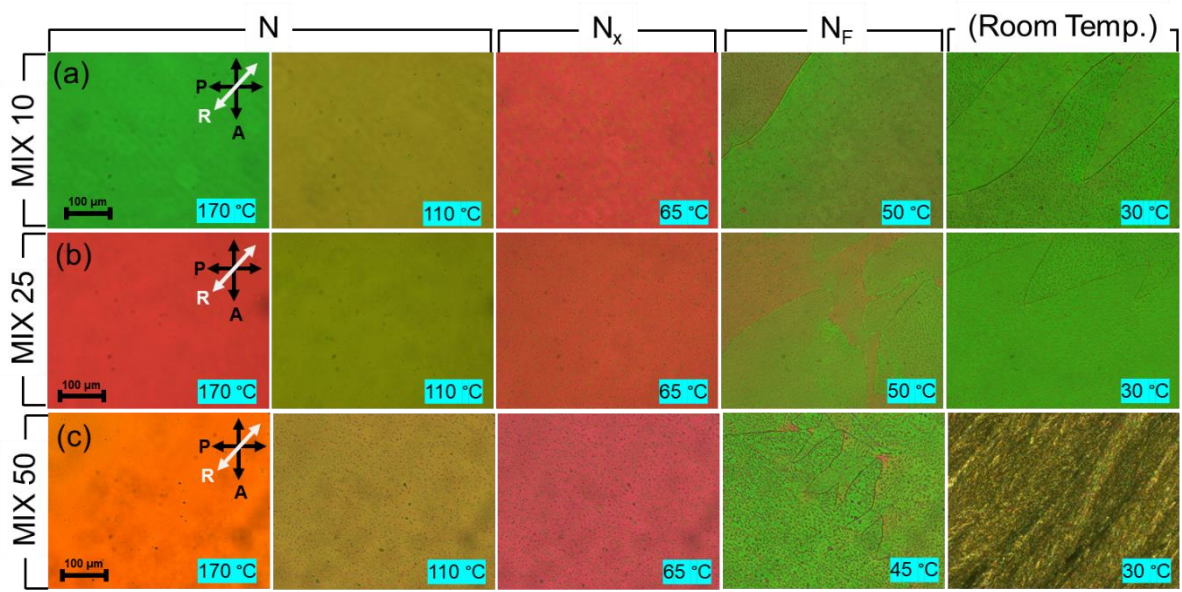

**Fig. 2 (a, b, c).** POM textures of 9 μm anti-parallel buffed planar aligned cells, filled with MIX 10, MIX 25, and MIX 50 respectively, acquired at different temperatures. The analyzer (A) and polarizer (P) orientations are indicated. The white arrow (R) denotes the rubbing direction. The scale bar represents a length of 100 μm

On cooling from the isotropic phase, all mixtures form perfect, homogeneous planar textures. On further cooling down to the $N_F$ phase, the texture's homogeneity persists with only color variation due to the change in birefringence. However, in the $N_F$ phase all three mixtures including DIO (but not WJ-16) show two-domain texture of opposite chirality which can be observed between slightly uncrossed polarizers (Fig. 3). Such domains are characteristic to the $N_F$ phase, which appears in anti-parallel rubbed planar cells due to the polar azimuthal anchoring energy [26]. Crystalline phase formation was exclusively observed in MIX 50 at room temperature, the mixture with the highest WJ-16 concentration (50%). For other mixtures, crystallization was only observed after overnight cooling below the room temperature.

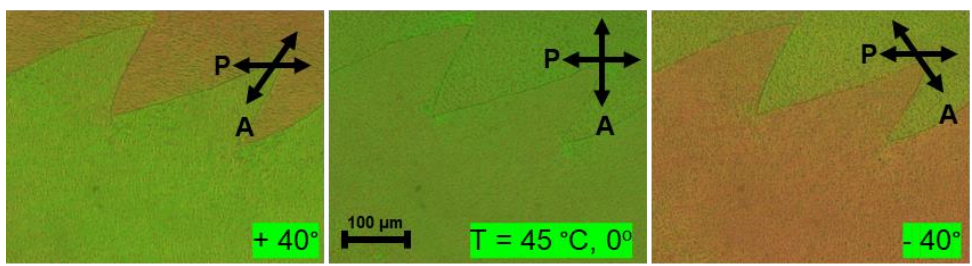

**Fig. 3.** POM textures of a 9 μm anti-parallel buffed planar cell, filled with MIX 25 at a temperature of 45 °C. Domains of the opposite chirality are observed upon rotation of the polarizer in opposite directions from the crossed polarizers position

In the homeotropically aligned cells, dark optical textures indicative of perfect homeotropic alignment were observed in the N and $N_x$ phases due to the minimal birefringence. In the $N_F$ phase, a more complex, grainy/sandy texture was observed, signifying the emergence of complex molecular ordering. In uncoated cells, disordered *Schlieren* textures were observed in the N and $N_x$ phases. In the $N_F$ phase, a mixture of *Schlieren* texture and complex domain structures were observed, resulting from an interaction of the polar ordering with the unaligned cell. The textures observed in both homeotropic and uncoated cells were consistent with those observed for the pure DIO compound

## 3.2 Birefringence

Birefringence ($\Delta n$) measurements were carried out employing an optical spectral technique [27] using a 9 µm homogenous planar-aligned cell. Transmittance ($T$) spectra were acquired using an Avantes AvaSpec-2048 fiber spectrometer with an achromatic light source, as a function of temperature, spanning from the Iso-N transition to the $N_F$ phase for all mixtures. The transmittance $T$ of a homogeneous planar-aligned cell is given by:

$$T = A \sin^2\left(\frac{\pi \cdot \Delta n(\lambda) \cdot d}{\lambda}\right) + B \qquad (1)$$

where $A$ is the amplitude factor, $B$ is the leakage offset of light through the cell, $d$ is the cell thickness, and $\Delta n(\lambda) = k \cdot \frac{\lambda^2 \cdot \lambda^{*2}}{\lambda^2 - \lambda^{*2}}$ is the birefringence dispersion, governed by the extended Cauchy equation. Birefringence data at a wavelength of 550 nm were calculated using a custom-developed software [27] and are presented in Fig. 4

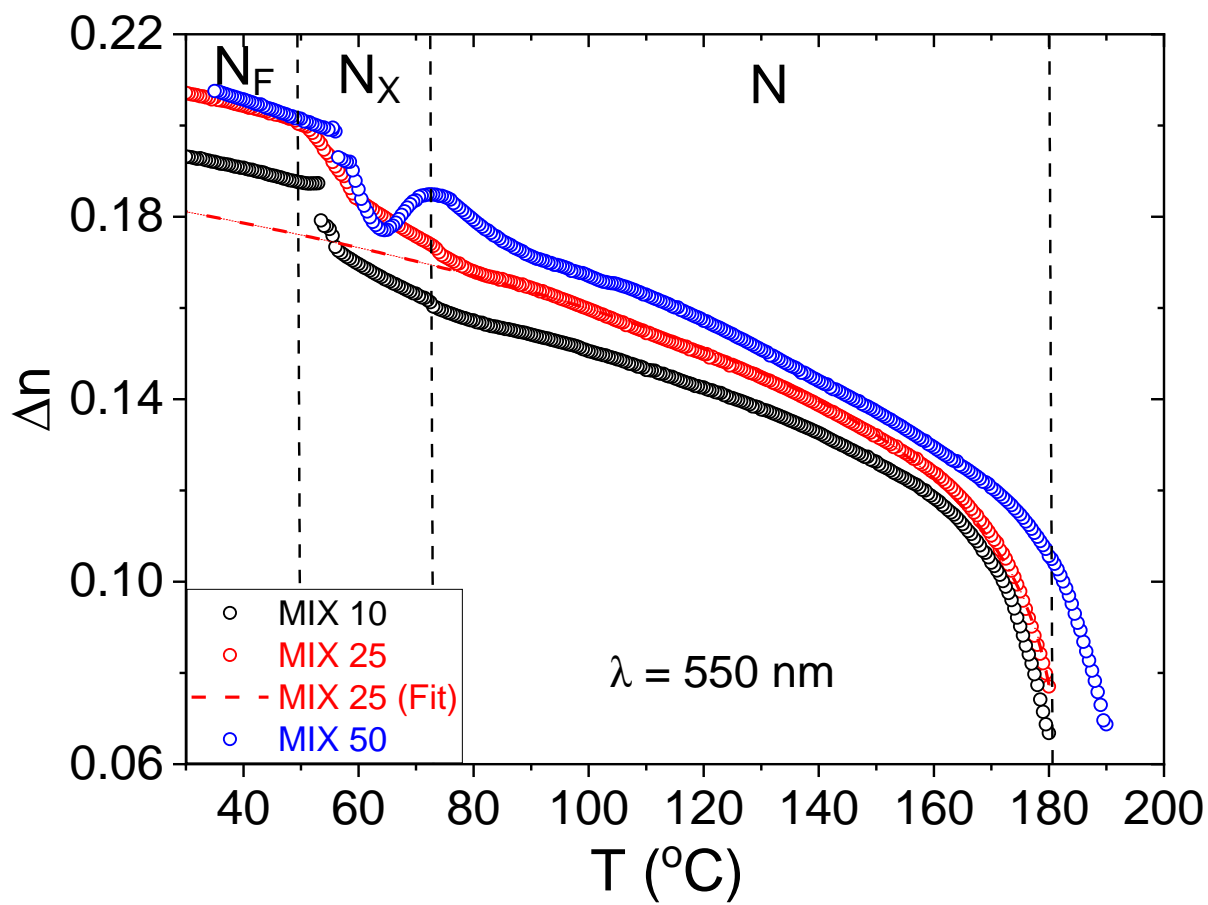

**Fig. 4**. The temperature dependence of birefringence ($\Delta n$) for MIX 10, MIX 25, and MIX 50 measured for a wavelength of $\lambda = 550$ nm using a 9 µm homogeneous planar-aligned cell. The experimental birefringence data fitted to Haller's equation (solid red line) for MIX 25 are shown as an example of the fitting accuracy. The vertical dotted lines show the phase transitions for MIX 25.

The temperature dependence of the birefringence in the high-temperature conventional N phase was fitted to the Haller equation, given as:

$$\Delta n(T) = \Delta n_o \cdot Q = \Delta n_o \cdot \left(1 - \frac{T}{T_{Iso-N}}\right)^\beta \qquad (2)$$

where $\Delta n_o$ is the maximum birefringence for the order parameter Q = 1, $T_{Iso-N}$ is the Isotropic to Nematic phase transition temperature, and β is the exponent, determined to be 0.2 ± 0.02, for an ordinary/conventional nematic phase. The best fit of experimental data to Eq. (2) confirms the conventional nematic behavior of the high-temperature N phase in the mixtures, with fitting parameters listed in Table 2

| System | $\Delta n_o$ | β | $T_{Iso-N}$ |
|---|---|---|---|
| MIX 10 | 0.18 | 0.19 | 179.0 |
| MIX 25 | 0.19 | 0.20 | 180.5 |
| MIX 50 | 0.19 | 0.22 | 190.0 |

**Table 2:** Fit parameters of Haller's equation for the experimental birefringence data of the DIO + WJ-16 mixtures

As shown in Fig. 4, the birefringence increases gradually with decreasing temperature in the N phase. Upon transition from the N phase, the birefringence curve deviates from the Haller's equation, exhibiting an increased value in the $N_x$ phase on cooling. Subsequently, a discontinuous jump in the birefringence is observed at the $N_x$-$N_F$ transition temperature. The birefringence curves of MIX 10 and MIX 25 are similar to DIO. For MIX 50, a dip in the birefringence curve is observed around 60 °C. However, no change in the optical textures was observed (see Fig. 2, MIX 50, T = 45 °C). The observed dip in the birefringence curve is due to the co-existence of both polar and non-polar domains and defects in the intermediate $N_x$ region, indicative of competing phase behaviors due to equal proportions of the two distinct compounds [28]

*3.3. Electrical Measurements*

A ferroelectric nematic ($N_F$) phase is characterized by the presence of a macroscopic spontaneous polarization (Ps). Here, we examine the ferroelectric behavior of mixtures by using the conventional reversal switching current technique [29]. An external electric field of suitable waveform (triangular), generated from an Agilent 33120A signal generator and amplified using a high-voltage amplifier (TReK PZD700), was applied across the liquid crystal cell. The resulting response across a 1 kΩ resistive load was monitored using a digital oscilloscope

Fig. 5 shows the switching-current response of a 4 μm planar cell under an applied triangular wave ($V_{pp}$ = 24 V, $f$ = 20 Hz) for (a) WJ-16, (b) DIO, and (c) MIX 25 in the ferroelectric ($N_F$) and non-ferroelectric (N) phases, respectively. WJ-16 shows no spontaneous polarization switching peak in the full temperature range, including the N and SmA phase, confirming that this sample is not ferroelectric. For DIO in the $N_F$ phase, a sharp, well-defined peak in the current transient confirms the existence of spontaneous polarization, whereas in the N phase, the absence of a current peak confirms the paraelectric behavior

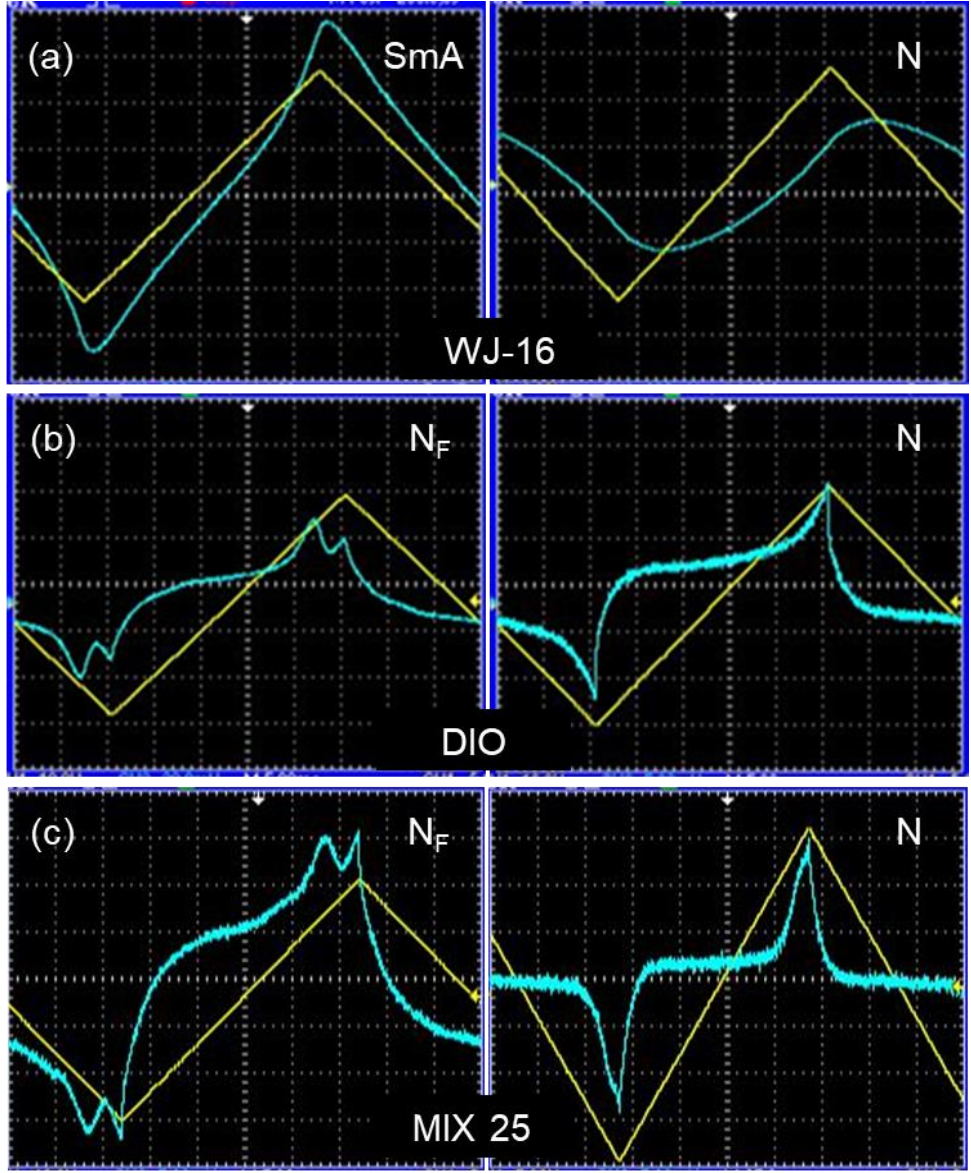

**Fig. 5**. Switching-current response of a 4 μm homogeneous planar-aligned indium tin oxide (ITO) cell under an applied triangular wave ($V_{pp}$ = 24 V, $f$ = 20 Hz) for (a) WJ-16, (b) DIO, and (c) MIX 25 in the ferroelectric ($N_F$) and non-ferroelectric (SmA, N) phases respectively

All three binary mixtures show a transient response similar to DIO. The response for MIX 25 (Fig. 5 (c)) is shown as an example. The integral of the current peak gives the value of Ps for MIX 25 to be 2.4 μC/cm$^2$

## 3.4. Dielectric Spectroscopy

### 3.4.1. Motivation and Methodology

Dielectric spectroscopy is one of the most sensitive techniques for the study of ferroelectric and other polar materials/phases. This was successfully employed for the characterization of ferro- [30,31,32,33] / antiferro- [34,35] and ferri-electric [36,37] liquid crystalline phases. One of the most important parameters of anisotropic media, such as liquid crystals, is the dielectric anisotropy, $\Delta\varepsilon_a = \Delta\varepsilon_\parallel - \Delta\varepsilon_\perp$, which governs the electro-optic switching and the Freedericksz transition. To obtain the value of the dielectric anisotropy, it is necessary to measure the real values of the perpendicular component of dielectric permittivity, $\Delta\varepsilon_\perp$, in planar cells and parallel component of dielectric permittivity, $\Delta\varepsilon_\parallel$, in homeotropic cells. Dielectric spectroscopy measurements over a frequency range of 0.01 Hz – 10 MHz were made using a broadband Alpha High Resolution Dielectric Analyzer (Novocontrol GmbH, Germany). The commercial cells of 4 μm and 9 μm thicknesses with both homeotropic and planar alignment were used. These commercial cells use ITO electrodes with low sheet resistance (10 Ω/□), which moves the frequency of parasitic ITO peak (arising from the sheet resistance of ITO in series with the capacitance of the cell) outside the measured frequency range. The measurements were carried out under the application of weak voltage of 0.1 V, applied across the cell. The temperature of the sample is stabilized to within ±0.05 ºC

*3.4.2. Dielectric Spectroscopy of homeotropic cells*

We started with the measurement of the parallel component of dielectric permittivity. The phase sequences of all three mixtures are similar to DIO (Table 1), therefore, we present the dielectric spectroscopy results for a MIX 25 only. Fig. 6 shows the temperature dependencies of the total dielectric permittivity (i.e., real (ε') and imaginary (ε'') parts of complex permittivity) for 9 μm homeotropic cell of MIX 25.

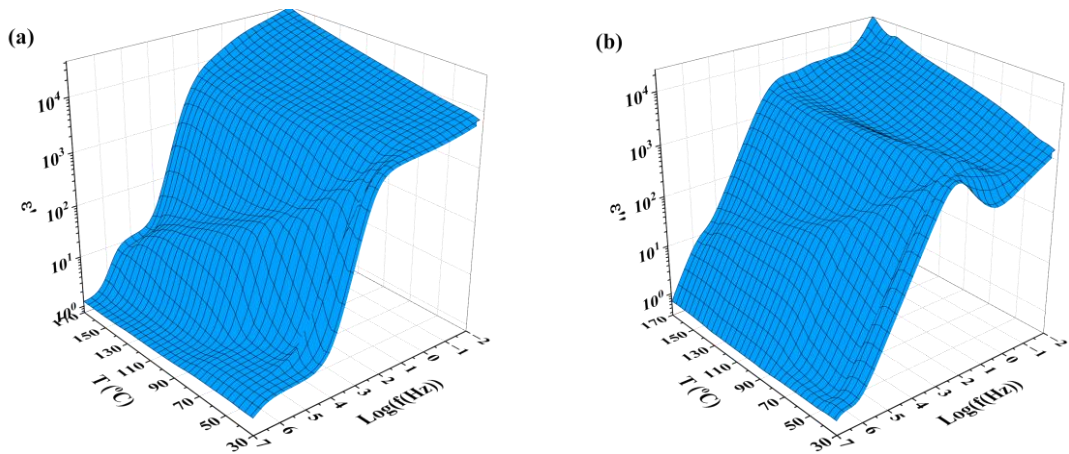

**Fig. 6.** Temperature dependence of dielectric spectra of (a) permittivity ε' and (b) loss ε'' measured in 9 μm MIX 25 *homeotropic* cell.

Fig. 7 (a) shows 2-D dielectric loss spectra at different temperatures for the 3-D plot of dielectric loss as shown in Fig. 6 (b). The dielectric spectra were analyzed using the Novocontrol WINDETA program. The complex permittivity data were fitted to the Havriliak-Negami Eq., as given below, where the DC conductivity term is also included [38]:

$$\varepsilon^* = -\frac{i\sigma}{\varepsilon_0 \omega} + \varepsilon_\infty + \sum_{j=0}^{n} \frac{\Delta\varepsilon_j}{\left[1+(i\omega\tau_j)^{\alpha_j}\right]^{\beta_j}} \tag{3}$$

Here, $\varepsilon^*$ is the complex permittivity, and $\varepsilon_\infty$ is the high-frequency permittivity. The latter includes electronic and atomic polarizabilities of the material. $\omega$ is the angular frequency of the probe field, $\varepsilon_o$ is the permittivity of free space, $\sigma$ is the DC conductivity, $\tau_j$ is the relaxation time, $\Delta\varepsilon_j$ is the dielectric strength of the $j^{th}$ relaxation process, $\alpha_j$ and $\beta_j$ are the corresponding symmetric and the asymmetric broadening parameters of the distribution of relaxation times

Fig. 7 (b-d) is an example of the fitting of the dielectric loss spectra to three relaxation processes at three different temperatures: 80 °C, 52 °C, 35 °C

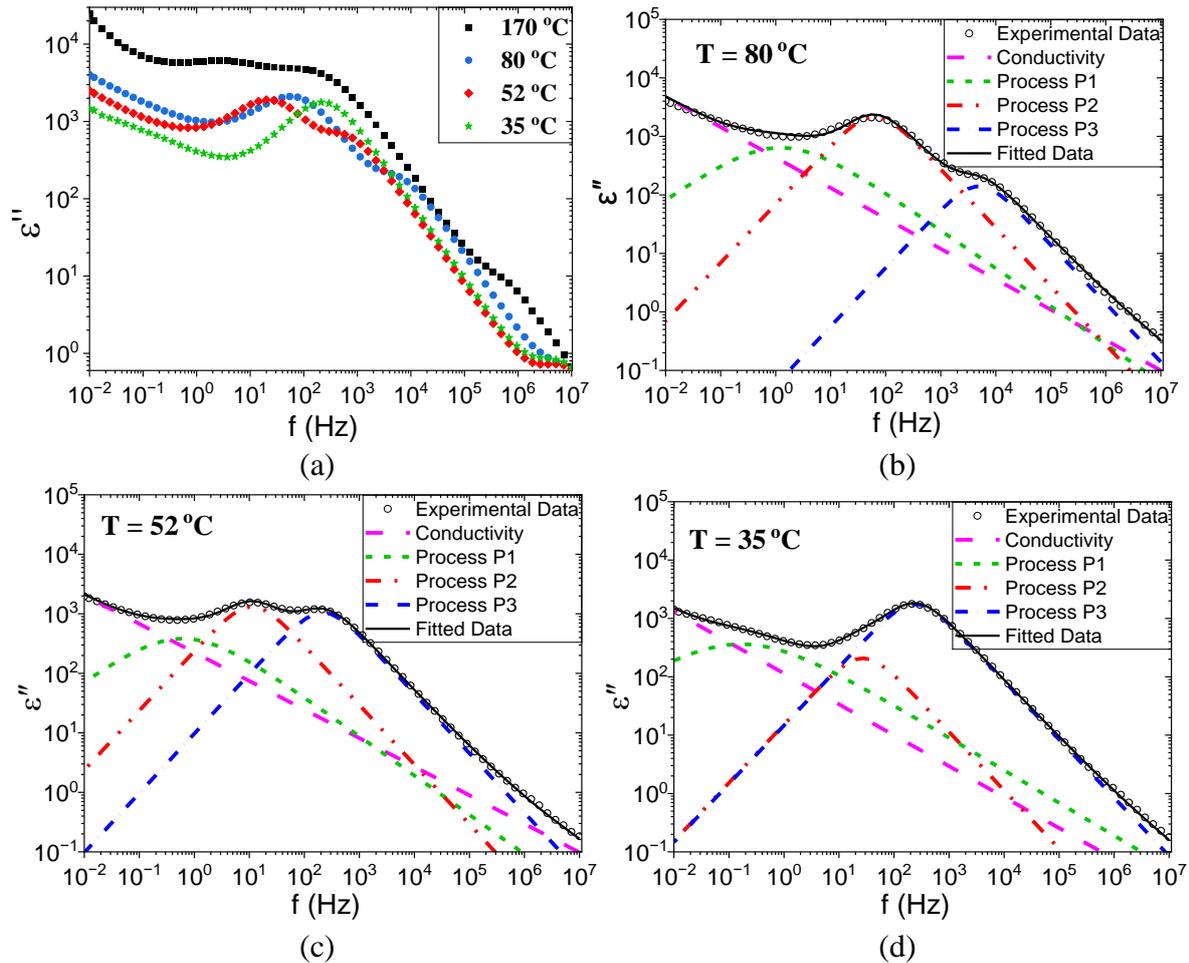

(a)

(b)

(c)

(d)

**Fig. 7** (a) The frequency dependence of the imaginary part of permittivity (ε") at different temperatures; (b-d) shows the examples of fitting of the imaginary part of permittivity to Eq. 3 at different temperatures

There are three relaxation processes observed in the dielectric spectra of mixtures termed as P1 - P3, with an increase in the relaxation frequency. The dielectric spectra were fitted to Eq. (3) with fixed stretching parameters (α, β), and from the fitting, we obtain β = 1 for all processes. The symmetric stretching parameter α = 1 for both processes P2 and P3, implying that these relaxation processes are pure Debye type. The symmetric stretching parameter for process P1, $α_1$, lies in the range 0.65 - 0.9 depending on the temperature, implying that this is not a molecular relaxation process

The physical nature/origin of the relaxation processes P1-P3 was defined in Ref. [23, 39, 40]. The lowest frequency relaxation process P1 is assigned to the space charge/interfacial polarization produced by the mobility of ions and, finally, their accumulation onto the electrodes. The process P2 can be assigned to the flip-flop mode, and the process P3 corresponds to the rotations around the long molecular axis, as reported in the literature [39,40,41]

Fig. 8 shows the temperature dependencies of the dielectric strengths and relaxation frequencies of three relaxation processes, P1-P3 and their sum for 9 μm homeotropic cell deduced from the fitting of dielectric spectra shown in Figs. 6-7

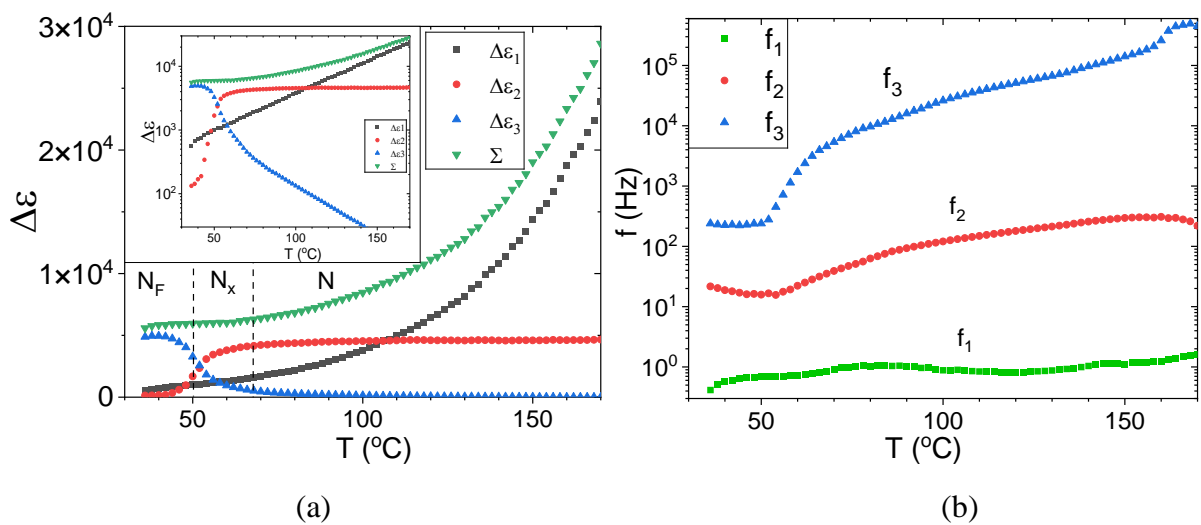

(a)  (b)

(b)

**Fig. 8.** Temperature dependence of (a) dielectric permittivities and (b) relaxation frequencies of MIX 25 in 9 μm *homeotropic* cell.

Upon examining Fig. 8 (a), the observed temperature dependencies of all three relaxation processes, P1-P3, immediately capture one's attention. The dielectric strength of ionic relaxation process P1 ($\Delta\varepsilon_1$) increases exponentially on heating, which appears as a straight line on the log Y-axis (see the inset). The dielectric strength of the second relaxation process P2 ($\Delta\varepsilon_2$) in non-ferroelectric phases exhibits colossal permittivity CP (~ 4500), is independent of the temperature, and is identical to the second relaxation process (P2) observed in compound WJ-16 [23] but not observed in DIO. The temperature dependence of dielectric strength of the fastest relaxation process P3 ($\Delta\varepsilon_3$) is typical for paraelectric-ferroelectric phase transition, similar to P2 in DIO. In the ordinary non-ferroelectric N phases, typical exponential growth is exhibited on cooling, and then in the $N_X$ phase the Curie-Weiss law is followed at the transition to the ferroelectric phase, similar to P2 in DIO. The temperature dependence of the dielectric strength of P3 shows no discontinuity at the N-Nx phase transition, and both phases dielectrically appear as a single phase. However, it is noted that the Nx phase was defined to be different from the N phase. Originally observed for DIO, the Nx phase was found to be density modulated antiferroelectric, $SmZ_A$ phase [42], or just "splay nematic", $N_S$' [43]. Since then, this Nx phase has been reported in in a range of additional compounds as well [41,44,45]. Dielectrically the $N_X$-$N_F$ phase transition is of second order, while N-$N_F$ is the first order phase transition [46].

Summarizing these data and ignoring the parasitic ionic process, the horizontal component of dielectric permittivity in the ferroelectric $N_F$ phase, $\Delta\varepsilon_\parallel \approx 5000$, is due to the Goldstone mode and in ordinary N phase, $\Delta\varepsilon_\parallel \approx 4500$, is due to the SPE mode [23]

*3.4.3. Dielectric Spectroscopy of Planar Cells*

The perpendicular component of dielectric permittivity $\Delta\varepsilon_\perp$ was measured in 9 μm commercial planar cells. The obtained dielectric spectra were fitted to Eq. (3), similar to the homeotropic cells

Fig. 9 (a) shows the temperature dependencies of the total dielectric strengths and relaxation frequencies of two relaxation processes observed in 9 μm commercial planar cells. From the comparison of Fig. 8 (a) and Fig. 9 (a), one can find that the temperature dependencies of the dielectric processes are very different from each other. Firstly, in contrast to the homeotropic cell, the ionic process is not observed in the planar cell, and the other two processes correspond to the processes P2- P3. Secondly, and most importantly, the total dielectric permittivity of the planar cell is independent of temperature or, according to the arguments put forward by the Boulder group [9] is limited by the capacitance of alignment layers

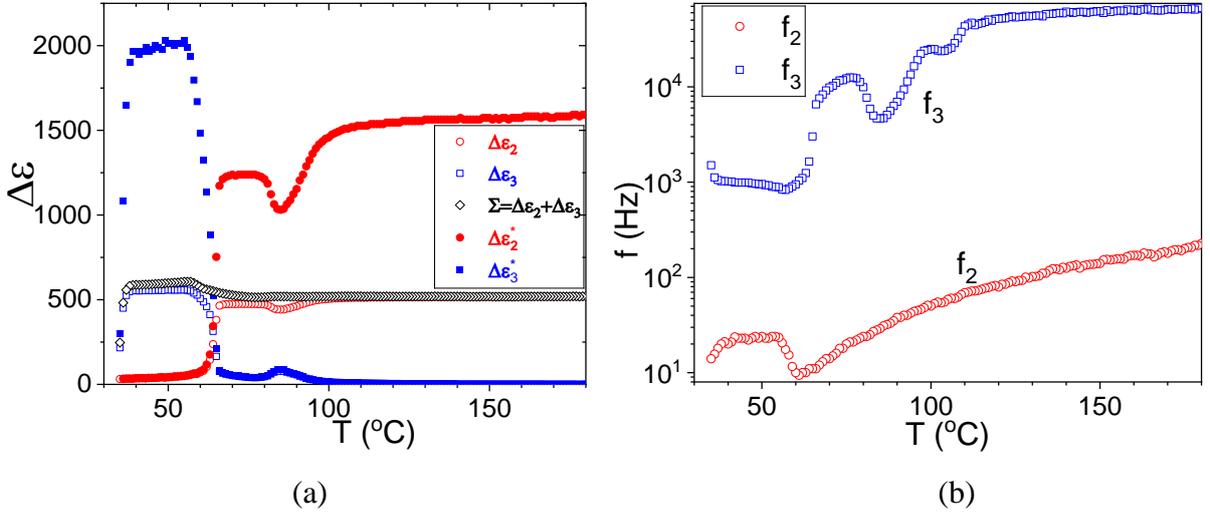

**Fig. 9.** Temperature dependence of (a) dielectric strengths showing the original (open symbols) data and recovered (filled symbols) dielectric strengths, and (b) relaxation frequencies of processes P2 and P3 for MIX 25 in 9 μm *planar* cell.

Recently, N. Clark have reported the effect of the insulating alignment layers on the apparent/measured values of the dielectric permittivity. He noticed that the apparent capacitance $C_{app}$ of an LC cell is a combination of three capacitances in series: the capacitance of the LC layer $C_{LC}$ and the two capacitances of the insulating alignment layers, $C_{al}$. Hence, the total (apparent) capacitance is $C_{app} = \frac{C_{LC} \cdot C_{al}}{C_{LC} + C_{al}}$. There can be two possible opposite cases: the ordinary case where $C_{LC} \ll C_{al}$, and the extraordinary case, where $C_{LC} \gg C_{al}$. In the ordinary one, i.e., in materials with low/moderate dielectric permittivity, the capacitance of the LC cell $C_{LC} \ll C_{al}$ and the apparent capacitance is the capacitance of LC cell, $C_{app} = C_{LC}$. Therefore, in the ordinary case, this gives a real value of capacitance and permittivity. However, in materials with very high dielectric permittivity ($\varepsilon > 10000$), such as in the ferroelectric $N_F$ phase, the capacitance of the LC cell can easily exceed the capacitance of the alignment layer, $C_{LC} \gg C_{al}$. In such a case, the apparent capacitance is limited by the capacitance of the insulating layers, $C_{al}$. We term it as "Clark's limit", $C_{app} \lesssim C_{al}$. This means that on approaching this limit, the apparent dielectric permittivity will show very weak temperature dependences as it can be observed in the planar cell; see Fig. 9 (a). Such a limit is also observed in other high ε materials such as bent-core LCs [47,48] and SPE nematics [23].

However, recently, we have shown several methods to recover the real value of capacitance/ dielectric permittivity from the limited apparent value [49]. Rewriting the previous equation for the capacitance of LC layers as:

$$C_{LC} = \frac{C_{al} \cdot C_{app}}{C_{al} - C_{app}} \qquad (4)$$

and knowing the value of the capacitance of the alignment layers, it is possible to restore the real capacitance from its apparent value

The Eq. (4) can be written in terms of dielectric permittivity as:

$$\varepsilon_{LC} = \frac{d_{LC} \cdot \varepsilon_{app} \cdot \varepsilon_{al}}{(d_{LC} \cdot \varepsilon_{al} + 2 d_{al} \cdot \varepsilon_{al} - d_{al} \cdot \varepsilon_{app})} \cong \frac{d_{LC} \cdot \varepsilon_{app} \cdot \varepsilon_{al}}{d_{LC} \cdot \varepsilon_{al} - d_{al} \cdot \varepsilon_{app}} \qquad (5)$$

and knowing the thickness and dielectric permittivity of alignment layers, it is possible to establish the real dielectric permittivity from its apparent value

Substituting $d_{al}$ = 20 nm and $\varepsilon_{al}$ = 3.8 (available in the data sheet of commercial cells) to Eq. (5), we obtain the correct values of dielectric permittivity with temperature dependence, as shown in Fig. 9 (a), as closed symbols. The recovered values of the perpendicular component of the dielectric permittivity are $\Delta\varepsilon_\perp \approx 2000$ in the ferroelectric $N_F$ phase and $\Delta\varepsilon_\perp \approx 1500$ in ordinary N phase. Interestingly, the value of the dielectric anisotropy are the same in both $N_F$ and N phases: $\Delta\varepsilon_a = \Delta\varepsilon_\parallel - \Delta\varepsilon_\perp = 5000 - 2000 = 4500 - 1500 = 3000$

*3.4.4. Discussion on Colossal Permittivity*

Now, it is the right time to return to different definitions of the paraelectric phase. Historically, the paraelectric is the non-ferroelectric phase, which shows on cooling the transition to its counterpart, the ferroelectric phase, described by the Curie-Weiss Law. However, according to the modern view, in the non-ferroelectric materials, the term "paraelectric" is often used instead of ordinary "dielectrics". We recently reported the paraelectric (according to the modern definition of paraelectrics) behavior of a non-ferroelectric compound (WJ-16), which exhibits colossal values of permittivity and hence was defined as superparaelectric [23]. However, the results obtained for the mixtures necessitate a revision of this assignment. According to Fig. 8 (a), in the ordinary nematic phase, we observed two different relaxation modes: the original paraelectric process as P3 and the independent CP process as P2. The process P3 is a paraelectric mode that follows the Curie - Weiss law and relates to the classical paraelectric- ferroelectric transition. The P2 mode exists rather independently along with the paraelectric P3. Taking into account that this temperature-independent mode also exists in the SmA and even in isotropic phases, it cannot be related to paraelectric and/or superparaelectric. Therefore, strictly speaking, the P2 process cannot be called SPE as it was assigned for pure WJ-16 compound [23]

According to the modern view, high dielectric permittivity reflects an extremely strong collective molecular dipolar contribution. In solid SPE materials, this is a contribution from well-aligned molecular clusters [50]. The dielectric permittivity depends on the size of ferroelectric domains/clusters, which is temperature-dependent but does also depend on the size of the sample. In other words, large values of the dielectric permittivity imply a dipolar orientation of the polar clusters. Such behavior is reported in bent-core LCs [47,48], which also consist of polar/ferroelectric cybotactic clusters that, in turn, grow upon cooling. In our mixtures and pure WJ-16, such clusters were not observed, and the dielectric strength is independent of frequency but linearly proportional to the cell thickness [23]. This is indeed an argument why P2 in the mixtures cannot be SPE. Such linear dependence on the cell thickness is observed in the $N_F$ phase [8,9,10] and other systems such as surface-stabilized ferroelectric SmC* [33] and bent-core LCs [51]. This implies that liquid crystals have a long-range directional order and the correlation length is limited by the cell thickness. In such systems, the molecular dynamics (phason mode) can be modeled with boundary conditions at the electrodes, which gives a linear dependence of dielectric strength on the cell thickness [8,9,10,33]

While the dielectric permittivity in $N_F$ and in our sample linearly depends on the cell thickness, they have different physical origins: the phason mode in the NF and the amplitude mode in SPE. In the NF phase, the assignment as phason mode is due to a collective fluctuation of the director and polarization. In WJ-16 and in the ordinary nematic phase of the mixtures reported here, the absence of spontaneous polarization explicitly excludes the phason mode as the origin of the relaxation process P2. Consequently, P2 must be attributed to an amplitude mode, reflecting induced polarization as a response to the electric field. The relaxation process P2 is associated with CP and similar linear dependence on thickness but they have different physical origin. This cannot be a phason mode due to the absence of spontaneous polarization; hence, this should be an amplitude mode. However, the effect is different from the classical amplitude mode with a para- to ferroelectric transition (P3, Fig. 8 (a)). The physical origin of this process is still not clear and requires further study but it is related to the long-range character of the dipole-dipole interaction). However, taking into account that it is not para-/ferroelectric process but exhibits CP, it can be assumed as a High-Permittivity dielectric mode or, briefly and literally, "HiPer Dielectrics" or simply "Hyper Dielectrics" to avoid possible confusion with spelling

## 4. Conclusion

In this paper, we prepared and studied binary mixtures of a well-known ferronematic DIO with the recently reported non-ferroelectric compound WJ-16, which shows colossal permittivity and superparaelectricity (SPE). We investigated three mixtures with different content of WJ-16 as 10, 25, 50% in DIO host. Our original expectation was the development of new nematic materials with both ferroelectric and non-ferroelectric CP phases. The prepared mixtures were studied by using different techniques such as POM textures, broadband dielectric spectroscopy, electrical switching study, etc. The results confirm the presence of a low temperature ferroelectric phase in all three mixtures, with a phase sequence similar to DIO. Interestingly, the non-ferroelectric phases in mixtures exhibit a CP mode. This CP mode was originally observed in WJ-16 and was termed as superparaelectric. However, the dielectric spectroscopy of mixtures shows two distinct relaxation processes: the typical paraelectric response and additionally the CP mode. Therefore, in these mixtures this CP mode cannot be identified as a SPE and needs to be re-defined as a Hyper Dielectric mode. This is the first direct demonstration of materials with both ferroelectric and Hyper-dielectric phases in liquid crystalline materials. Moreover, it shows that high dipole LC materials enable the exploration of new condensed matter physics where fluidity, self-assembly and polarity lead to novel physical phenomena not observed for solid state materials. The hyper dielectric phase has a good potential as a working media for the supercapacitors industry.

**Acknowledgment**

The work in Dublin was funded by the US-Ireland, SFI 21/US/3788. WJ thanks the CSC, China for a PhD scholarship.